\documentclass[preprint,12pt]{elsarticle}

\usepackage{amssymb}

\journal{Computer Physics Communications}

\usepackage{slashbox}
\usepackage[boxed]{algorithm2e}
\usepackage{url}

\def\epspdffile#1{
  \scalebox{0.32}{\includegraphics[width=30cm]{#1.eps}}
}

\def\nvidia{NVIDIA}
\renewcommand{\today}{November 14, 2009}

\begin{document}

\begin{frontmatter}




\title{Solving Lattice QCD systems of equations using mixed precision solvers on GPUs}


\author[cfa,iic]{M.~A.~Clark}
\author[ccs,buphy]{R.~Babich}
\author[nwesam,nwmse]{K.~Barros}
\author[ccs,buphy]{R.~C.~Brower}
\author[ccs,buphy]{C.~Rebbi}

\address[cfa]{Harvard-Smithsonian Center for Astrophysics, \\60 Garden St, Cambridge,  MA 02138, USA}
\address[iic]{Initiative in Innovative Computing, Harvard University School of Engineering and Applied Sciences, 29 Oxford St, Cambridge,  MA 02138, USA}
\address[ccs]{Center for Computational Science, Boston University, \\3 Cummington St, Boston, MA 02215, USA}
\address[buphy]{Physics Department, Boston University, \\590 Commonwealth Avenue, Boston, MA 02215, USA}
\address[nwesam]{Department of Engineering Sciences and Applied Mathematics, Northwestern University, 2220 Campus Drive, Evanston, IL 60208, USA}
\address[nwmse]{Department of Materials Science and Engineering, Northwestern
  University, \\2220 Campus Drive, Evanston, IL 60208, USA}

\begin{abstract}
  Modern graphics hardware is designed for highly parallel numerical
  tasks and promises significant cost and performance benefits for
  many scientific applications.  One such application is lattice
  quantum chromodyamics (lattice QCD), where the main computational
  challenge is to efficiently solve the discretized Dirac equation in
  the presence of an $SU(3)$ gauge field.  Using \nvidia's CUDA
  platform we have implemented a Wilson-Dirac sparse matrix-vector
  product that performs at up to 40 Gflops, 135 Gflops and 212 Gflops
  for double, single and half precision respectively on \nvidia's
  GeForce GTX 280 GPU.  We have developed a new mixed precision
  approach for Krylov solvers using {\it reliable updates} which
  allows for full double precision accuracy while using only single or
  half precision arithmetic for the bulk of the computation.  The
  resulting BiCGstab and CG solvers run in excess of 100 Gflops and,
  in terms of iterations until convergence, perform better than the
  usual defect-correction approach for mixed precision.
\end{abstract}


\end{frontmatter}


\section{Introduction}
For decades, Moore's law has reliably given a doubling of the number
of transistors per chip about every 18 months, a trend that continues
to this day. In the past, such increases translated directly into
improved performance for serial code through higher clock rates,
larger caches, and increased exploitation of instruction-level
parallelism. Recently, however, such improvements have yielded
diminishing returns, bringing us to the era of multi-core CPUs. For
the intrinsically parallel tasks commonly found in scientific
computing, this is a welcome development.  Still, it is not obvious
that commodity processors, whose high clock rates and large caches
come at the expense of greater numbers of cores, represent the optimal
balance for highly parallel workloads. Graphics processing units
(GPUs), driven by the large video game market, represent a different
set of trade-offs. GPUs emphasize very high parallelism and memory
bandwidth, a recipe for astounding performance in many scientific
applications.

Lattice QCD (quantum chromodynamics) is the lattice discretized theory
of the strong force, that which binds together quarks in the nucleon.
In lattice QCD, the propagation of quarks is given by the inverse of
the Dirac operator, which is a large sparse matrix.  Hence, many
systems of linear equations must be solved involving this matrix; it
is this requirement that makes lattice QCD a grand challenge subject.
In such linear solvers, the application of the Dirac operator to a
vector is the most compute intensive kernel.  This work is an initial
exploration of how best to implement these solvers on a single GPU.

As recently as a few years ago, utilizing GPUs for general-purpose
computing required one to manipulate graphics primitives via APIs such
as OpenGL and associated shader languages.  A pioneering study in this
vein was presented in~\cite{Egri:2006zm}, where Dirac solvers were
shown to map well onto GPU architectures despite the limitations of
the programming model.  Our implementation relies on \nvidia's Compute
Unified Device Architecture (CUDA), embodied in the last two
generations of {\nvidia} GPUs, and the ``C for CUDA'' programming
language~\cite{Nvidia:2009}.  C for CUDA is a C-like language that
provides direct and relatively low-level access to the GPU via a
convenient software development toolkit and runtime environment.  This
work builds upon our initial investigation into mapping lattice QCD
onto CUDA~\cite{Barros:2008rd}.

A characteristic of current generation GPUs, the Cell processor and
even Intel's SSE vector units is that the performance of single
precision arithmetic is superior to that of double precision.  On CPUs
this difference in performance is usually only a factor of two,
whereas on GPUs the difference can be as much as an order of magnitude
(if double precision is supported at all).  Thus, strategic use of
precision in a GPU calculation is vitally important to obtaining high
performance.

Previous work using mixed precision to solve systems of linear
equations on GPUs have focused on defect-correction approaches:
calculate the residual in double precision, find the solution using
single precision, accumulate the solution in double precision and
repeat this process until convergence~\cite{Goddeke:2005}.  Thus most
operations are performed in single precision, but the final result is
accurate to double precision.  The disadvantage of defect-correction,
however, is that the Krylov search space that is built up is discarded
each time the solver is restarted.  Thus the total number of
iterations required can drastically increase compared to a standard
full double precision solve.  In this work we introduce a new method
for using mixed precision in the context of Krylov solvers,
repurposing the \emph{reliable updates} scheme of
\cite{Sleijpen:1996}.  Using a strategy whereby the
iterated residual is replaced periodically with the true residual
calculated at high precision, together with high precision groupwise
updates for the solution, we demonstrate a mixed precision approach
that does not require an explicit restart.  We shall show that this
performs better than defect-correction.

The paper is organized as follows: in \S \ref{sec:gpu} we give an
overview of GPU hardware and the CUDA programming model; \S
\ref{sec:wilson} introduces the Wilson-Dirac matrix; \S
\ref{sec:implement} describes how we have implemented an optimized
matrix-vector kernel; in \S \ref{sec:inverter} we describe our mixed
precision Krylov solvers and we end with some general conclusions in \S
\ref{sec:conclusion}.

\section{Graphics Processing Units}
\label{sec:gpu}
\subsection{Hardware}
\label{sec:hardware}
In this work we utilize {\nvidia} graphics cards as our testbed since
these support a mature API for general-purpose computing, described in
more detail below.  Many of the strategies we discuss will carry over
to GPUs produced by other manufacturers (e.g.,~AMD, Intel) as these
begin to support the emerging OpenCL standard~\cite{opencl}, which
accommodates a very similar programming model.  {\nvidia} markets three
lines of graphics cards. The GeForce series serves the lucrative
consumer video game market, while the Tesla series targets the high
performance computing (HPC) market. Tesla cards retain the core
architecture of the consumer cards but offer more device memory and
greater reliability, at the expense of lower memory bandwidth, no
video output and increased cost.  Finally, {\nvidia} markets the Quadro
line for professional graphics applications.

\begin{table}
\begin{tabular}{|l|c|c|c|c|c|} \hline
 &  &  & \multicolumn{2}{|c|}{Gflops} &  GiB \\ \hline 
Card & Cores & Bandwidth & 32-bit & 64-bit & Device RAM  \\ \hline
GeForce 8800 GTX & 128 & 86.4 & 518 & -  & 0.75 \\ \hline
Tesla C870            & 128 & 76.8 & 518 & -  & 1.5 \\ \hline
GeForce GTX 280   & 240 & 142  & 933 & 78 & 1.0 \\ \hline
Tesla C1060          & 240 & 102  & 933 & 78 & 4.0 \\ \hline
\end{tabular}
\caption{\label{table:specs}Specifications of representative {\nvidia} graphics cards.}
\end{table}

\begin{figure}[htb]
\begin{center}
\epspdffile{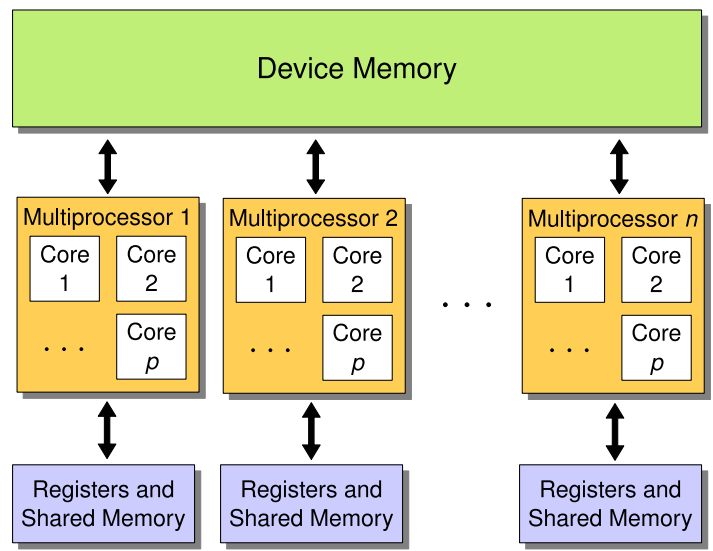}
\caption{\label{fig:arch}Architecture of a modern {\nvidia} graphics
  card.  In \nvidia's nomenclature, cores called {\it stream
    processors} (or {\it scalar processors}), and in current GPUs each
  multiprocessor has eight such cores.}
\end{center}
\end{figure}

To date, there have been roughly two generations of CUDA-enabled
GPUs. The flagship consumer cards of the previous and current
generation are the GeForce 8800 GTX and the GTX 280,\footnote{For the
  performance results reported below, we utilize a GeForce GTX 280.
  The more recently released GTX 285 is a higher-clocked variant that
  may be expected to yield 10-15 percent better performance.}
paralleled in the HPC line by the Tesla C870 and C1060. See Table
\ref{table:specs} for detailed specifications.  In the current generation,
double precision arithmetic is supported natively for the first time.

A modern {\nvidia} GPU contains many multiprocessors, each composed of
several cores, as illustrated in Figure \ref{fig:arch}. For example,
the GPU in the GeForce GTX 280 contains 30 multiprocessors and a total
of 240 cores. Primary storage on the card is provided by device
memory, which is shared among all multiprocessors and has a relatively
high latency. However, this latency can often be hidden with a high
multiprocessor occupancy: many active threads simultaneously loaded
and ready to execute.  An additional important consideration is that
the highest bandwidth from device memory is achieved when accesses are
coalesced; this occurs when groups of 16 threads access a contiguous,
properly aligned memory region.

Unlike typical CPUs, the GPU does not contain a large memory
cache. Instead, each multiprocessor has a relatively small amount of
fast shared memory (16 KiB on current hardware), which is manually
managed by the programmer and shared between threads. Shared memory is
orders of magnitude faster than device memory, so access to the latter
must be minimized.

Registers serve the same function on a GPU as they do on a
CPU. Namely, they appear as explicitly labeled operands in machine
code instructions generated by the compiler. Every active thread on a
multiprocessor is allocated a sufficient number of private registers
to execute the CUDA kernel. Unlike shared memory, registers cannot be
shared between threads. Another limitation is that data stored in
registers cannot be organized into an array and dynamically indexed.
\nvidia's previous generation GPUs provide 8,192 single precision
registers, while in the more recent generation the number has been
doubled to 16,384.
 
Each multiprocessor also provides two small read-only caches (8 KiB
each on current hardware): a texture cache and a constant cache.  The
former is optimized for reading data from global device memory that
has \(2d\) locality.  In addition, any data read through this cache
can be interpolated and scaled in hardware; these texture operations
are in addition to the raw flops listed in Table \ref{table:specs}.
While the interpolation capability is not of practical use for this
application, we do take advantage of scaling operations in our half
precision implementation (see \S \ref{sec:half} below).

Lastly there is the constant memory. This is 64 KiB of read only
storage on the GPU that is cached onto each multiprocessor.  The
values in constant memory must be set from code executing on the
device, and because of the constant cache's small size, it is mainly
useful only for reading parameters, numerical constants and perhaps
storing lookup tables.

No discussion regarding GPU hardware in the context of general purpose
computing is complete without mentioning the bottleneck of getting
data to and from the GPU through the PCI Express bus.  The typical
bandwidth of this bus, 6 GiBs\(^{-1}\), can be a severe constraint.  In
this application we implemented the entire solver on the GPU to reduce
the CPU-GPU data transfer.  The time required for the initial download of
the matrix and source vector, and the final upload of the solution
vector, is negligible compared to the operation time of the solver.

\subsection{The CUDA Programming Model}
\label{sec:cuda}

The CUDA platform\footnote{In the remainder of the paper, we
  colloquially use ``CUDA'' to refer not only to \nvidia's hardware
  architecture but also to the accompanying software stack and the C
  for CUDA language in particular.} provides direct access to the GPU
through a C-like programming language with minimal extensions
\cite{Nvidia:2009}. The CUDA platform includes a compiler that targets
the GPU, as well as a hardware driver and a runtime library. Higher
level libraries are also provided, including optimized BLAS and FFT
implementations.

A CUDA application works by spawning a very large number of threads,
as many as tens of thousands at once, which execute in parallel. For
example, in a lattice QCD application one might assign one thread to
each lattice site. The user specifies the organization and shared
memory usage of the threads when a CUDA kernel is invoked. As an
example, consider the CUDA code
\begin{verbatim}
dslashKernel<<<gridDim, blockDim, sharedBytes>>>(args); 
\end{verbatim}
which invokes \texttt{dslashKernel(args)} for execution by many
individual threads on the GPU.  Threads are grouped into thread
blocks, and the entire collection of thread blocks is called a
grid. The statement above tells the GPU to launch a kernel using gridDim
blocks, each containing blockDim threads. The compiler is instructed
to allocate sharedBytes bytes of shared memory to each block. This
shared memory allows for rapid communication between threads within a
thread block. CUDA provides primitives to allow synchronization
between threads within a thread block. However, no synchronization is
possible between different thread blocks (within a single kernel
invocation).

The GPU will dynamically schedule the thread blocks for execution.
The penalty for high latency operations can be reduced or eliminated
when there is a high multiprocessor occupancy: each multiprocessor
should have many threads simultaneously loaded and waiting for
execution. The challenges to achieving high multiprocessor occupancy
will be discussed in \S \ref{sec:local}. The GPU supports conditional
execution, but it is highly desirable that groups of 32 threads (a
thread warp) follow the same execution path. Otherwise, both execution
paths are serialized and executed by the entire warp.

\section{The Wilson-Dirac Matrix}
\label{sec:wilson}

The Wilson-Dirac matrix is a central difference discretization of the
Dirac operator, with the addition of a scaled Laplace matrix to remove
spurious fermion doublers.  When acting in a vector space that is the
tensor product of a 4-dimensional discretized Euclidean spacetime,
spin space and color space it is given by
\begin{eqnarray}
 M_{x,x'} & = & - \frac{1}{2} \displaystyle \sum_{\mu=1}^{4} \bigl(
(1-\gamma_\mu) U_x^\mu\, \delta_{x+\hat\mu,x'}\, +  
(1+\gamma_\mu)U_{x-\hat\mu}^{\mu \dagger}\, \delta_{x-\hat\mu,x'}\bigr) + 
(4 + m)\delta_{x,x'} \nonumber\\
& = & - \frac{1}{2} \displaystyle \sum_{\mu=1}^{4} \bigl(
 P^{-\mu} U_x^\mu\, \delta_{x+\hat\mu,x'}\, +  
P^{+\mu} U_{x-\hat\mu}^{\mu \dagger}\, \delta_{x-\hat\mu,x'}\bigr) + 
 (4 + m)\delta_{x,x'} \nonumber \\
& = &  - \frac{1}{2}D_{x,x'} + (4 + m) \delta_{x,x'}.
\end{eqnarray} 
Here \(\gamma_\mu\) are the \(4\times4\) Dirac matrices, which when
added to or subtracted from the identity form projectors \(P^{\pm\mu}\)
in {\it spin} space; \(U\) is the QCD gauge field which is a field of
SU(3) matrices acting in {\it color} space that live between the
spacetime sites (and hence are referred to as link matrices); and
\(m\) is the fermion mass parameter.  The indices \(x\) and \(x'\) are
spacetime indices (the spin and color indices have been suppressed for
brevity).  This matrix acts on a vector consisting of a complex-valued
12-component \emph{color-spinor} for each point in spacetime.  We
refer to the complete vector as a spinor field.

Quark physics require many solutions to systems of linear equations
involving this matrix; i.e., given the system of equations
\[
M\,x = b,
\]
we desire the solution vector \(x\), for many different source vectors
\(b\) and different gauge fields \(U\).  Given the sparsity and
dimensions of \(M\) (current state of the art calculations involve a
spacetime lattice with as many as \(64^3\times128\) color-spinors), it
is only feasible to consider iterative solvers for this problem.  In
this work we consider Krylov sub-space solvers.

If we apply an even-odd labelling to the lattice sites (also known as
red-black labelling) we see immediately that \(D_{x,x'}\) only
connects even sites to odd sites and vice versa.  It can be shown that
if the lattice sites are reordered according to this labelling, the
Schur complement of the Wilson-Dirac matrix has a condition number 2-3
times smaller than that of the full matrix~\cite{DeGrand:1990}.
Thus when solving such systems it is conventional to solve the Schur
complement system from which the solution to the original problem can
trivially be reconstructed.  If we choose to solve the Schur complement on the even
sites, the matrix is given by

\begin{equation}
  \hat{M}_{e,e} = 1_{ee} - \kappa^2 D_{eo}D_{oe},
\end{equation}
where \(D_{eo}\) (\(D_{oe}\)) represents the action of \(D_{x,x'}\)
connecting odd (even) to even (odd) sites only, and \(\kappa\) is
given by \(1 / (2 (4+m))\).  Using this form of the Wilson-Dirac matrix also
halves the vector memory requirements, which is extremely
advantageous, however, an extra temporary field is required when
applying the matrix if the original source is to be preserved.

Since the most compute intensive part of any Krylov solver is the
matrix-vector product, implementing an efficient GPU solver is
dependent on the application of \(D\) to a spinor field.  We shall herein
refer to the application of \(D\) to a spinor field as the {\it
  Dslash} operation.

Before continuing it is worth mentioning, if only to discount it, that
one option is to encode the Wilson-Dirac matrix explicitly as a sparse
matrix using one of the many packed sparse matrix formats (e.g.,
Compressesd Sparse Row (CSR), Packet (PKT), etc.) for which sparse
matrix-vector GPU libraries are available \cite{Bell:2008}.  On the
GTX 280, the library in \cite{Bell:2008} could achieve up to 30 Gflops
of sustained single precision performance for matrices with similar
structure as the one discussed here.  This is a very small percentage
of the peak performance (933 Gflops) of this card; this poor
performance is caused partly by the large ratio of peak floating point
performance to memory bandwidth.  However, as shall be shown below,
there are many symmetries of the Wilson-Dirac matrix that can be
used to reduce the required memory throughput.  Since such generic
libraries are by definition completely ignorant of such structure,
achieving high performance through the use of a generic library is not
possible.

\section{CUDA Implementation}
\label{sec:implement}

The discussion that follows is primarily focused on our single
precision implementation.  Specific issues related to double and half
precision are discussed in \S\ref{sec:double} and \S\ref{sec:half},
respectively.  All quoted performance results were obtained with
release 2.3 of the CUDA toolkit and driver version 190.29.

\subsection{Data Ordering}
\label{sec:data-order}
We consider a lattice of 4 dimensions (3 space and 1 time), splitting
the gauge and spinor fields into even and odd sub-lattices.  Our CUDA
implementation spawns one thread for each site in the (even or odd)
sub-lattice.  With 3 colors and 4 spin components, spinor fields
require 24 floats per lattice site. Gauge fields require 18 floats per
link. Following \cite{Egri:2006zm}, we employ a specialized data
layout. We do so because, as we have mentioned, maximum bandwidth is
obtained when 16 consecutive threads (a half warp) simultaneously read
16 primitive elements that are packed contiguously in device
memory. The available primitive elements include structures of 1, 2,
or 4 packed floats. Our testing indicates that, on current hardware,
the best bandwidth is achieved using float4 primitives. Spinors are
composed of 24 floats, so we use 6 arrays of float4s to store the
entire spinor field (stored within a single contiguous block of
memory). In this way, consecutive threads can simultaneously access
(nearly) consecutive elements from device memory. The gauge link
matrices are stored as either 8 or 12 floats (see Section
\ref{sec:bandwidth}), requiring 2 or 3 arrays of float4s respectively.

Since the lattice sites are split by parity, and because of boundary
effects, the half warp of 16 consecutive threads may access float4
objects that are nearly, but not exactly, contiguous in memory. The
texture cache mitigates the performance penalty of imperfect memory
accesses.

\subsection{Local storage requirements}
\label{sec:local}

Our tests indicate that it is desirable to have 192 active threads
per multiprocessor and that performance rapidly degrades with fewer
active threads. The multiprocessor occupancy is determined by the
register and shared memory requirements of the CUDA kernel.

Each thread in the Dslash operation must accumulate to a single output
spinor, which is composed of 24 floats and should reside in local
storage. In constructing the output spinor, the thread loops over all
neighboring sites. In each direction, a full spinor and gauge link
must be read. The neighboring spinor is immediately projected into a
half spinor and requires only 12 floats of local storage. The SU(3)
matrix representing the gauge link requires an additional 18
floats. Thus, at a minimum, 54 floats (plus any temporary registers
that the compiler deems necessary) are required per thread. If these
are to be stored entirely in the 16 KiB of shared memory, then at most
64 threads would be active on one multiprocessor.\footnote{The number
  of active threads must be a multiple of 32, the warp size.  A
  multiple of 64 is recommended.}  This number is much smaller than
our target, 192, and would negatively impact performance, and we
therefore use registers for additional data storage.  The GTX 280 has
16,384 registers per multiprocessor, providing 64 KiB of local
storage.  Using both shared memory and registers it is possible to
obtain 256 active threads per multiprocessor for the Dslash kernel.

For example, we store the SU(3) matrix elements in registers by
declaring
\begin{verbatim}
float g1,g2,g3,...,g18;
\end{verbatim}
We cannot use loops to express matrix operations on these
elements. Writing the full Dslash operation by hand, and without using
loops, would be tedious and error-prone. For this reason, we found it
expedient to automatically generate the lengthy Dslash CUDA code.  Our
current Dslash code generator was written in Python~\cite{Python}.

\subsection{Memory Bandwidth Requirements}
\label{sec:bandwidth}

The action of the Dslash kernel on a single site requires 1320
floating point operations, \(8(24+18)\) float loads and 24 float
saves.  This equates to a total of 1440 bytes, and thus performance is
severely limited by memory bandwidth.  It is therefore critical that
the calculation be structured in a manner that helps reduce the
quantity of transferred data.

Since the link matrices are elements of the \(SU(3)\) group, only 8
real numbers are required to parametrize the matrix.  The GPU
implementation in \cite{Egri:2006zm} used a simple 12 number
parametrization of the \(SU(3)\) matrix~\cite{De Forcrand:1986af}.  We
have investigated this approach, as well as a minimal 8 number parametrization
(see \ref{app:su3} for details).

Another trick to reduce the storage needed for the gauge field is to
exploit the gauge invariance of the Wilson-Dirac matrix.  In
particular we can impose a gauge transformation such that almost all
the gauge field links in the temporal direction are the unit matrix
(the exception being a surface term caused by the boundary
conditions).  Thus the gauge field need not be loaded when updating
the sites across the temporal direction.\footnote{For those readers
  without a quantum field theory background: for both the gauge
  transformation and the Dirac matrix basis change discussed below, we
  are imposing physically motivated similarity transformations upon
  the Wilson-Dirac matrix to increase its sparsity, and hence reduce
  the number of matrix elements that must be read in.}  An added
advantage of both gauge transforming and using an 8 number
representation is the reduced storage requirements, since device
memory is at such a premium on the GPU.

There are no parametrizations possible with the spinor field
components; however, it is possible to change the basis of the Dirac
matrices such that one of the four of these matrices is diagonal (we
choose \(\gamma_4\)).  In doing so, the spin projectors associated with
this matrix have only two non-zero elements, halving the number of
spinor components that must be read in (see \ref{app:gamma}).

Using all of these strategies together, i.e., with the 8 number
parametrization of the gauge field, gauge fixing applied and the
\(\gamma_4\) diagonalization, we (asymptotically) require only \(7(24)
+ 6(8)\) loads and 24 saves, for a total of 960 bytes.  Thus the
bandwidth requirements of the Dslash operation are reduced by a third.

\subsection{Single Precision Performance}

\begin{figure}[htb]
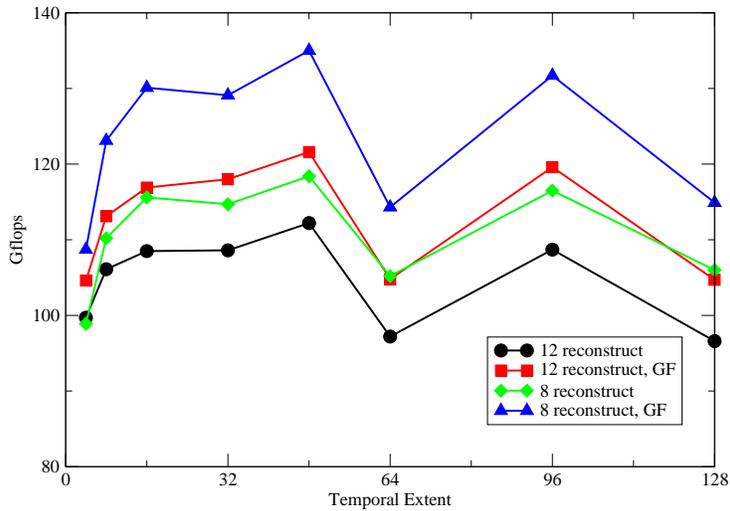

\begin{center}
\epspdffile{24_single}
 \caption{\label{fig:wilson-single-flops}Performance of the single
   precision even-odd preconditioned Wilson-Dirac matrix-vector
   product on a GTX 280 as a function of temporal length (spatial
   volume \(24^3\); GF denotes that temporal gauge fixing has been
   applied).}
\end{center}
\end{figure}
 
\begin{figure}[htb]
\begin{center}
\epspdffile{24_single_pc}
 \caption{\label{fig:wilson-single-pc-flops}Performance of the padded
   single precision even-odd preconditioned Wilson-Dirac matrix-vector
   product on a GTX 280 as a function of temporal length (spatial
   volume \(24^3\); GF denotes that temporal gauge fixing has been
   applied).}
\end{center}
\end{figure}
 
In Figure \ref{fig:wilson-single-flops}, we present single precision
performance results for a variety of different strategies for the
even-odd preconditioned Wilson-Dirac matrix on a range of different
volumes on the GTX 280.\footnote{In all cases, the reported
  performance numbers are ``effective Gflops'' that may be compared
  with implementations on traditional architectures. In particular,
  the nominal number of operations per lattice site does not include
  the extra work done in the SU(3) reconstruction, nor the savings
  associated with having trivial links in the time direction.}
Notable is the significant reduction in performance at \(T=64\) and
\(T=128\) due to {\it partition camping}.  This is where memory
conflicts occur when threads collide when reading from the device
memory.  Such conflicts can be overcome by padding the fields so that
the memory access patterns are not divisible by the partition width
(256 bytes on the GTX 280)~\cite{paulius}.  This is demonstrated in
Figure \ref{fig:wilson-single-pc-flops} where we again show the
performance as a function of temporal length, but now the contiguous
block of 6 sub-arrays of float4s with
\(\frac{1}{2}V=\frac{1}{2}N_XN_YN_ZN_T\) elements is padded such that
the beginning of each sub-array is separated by
\(\frac{1}{2}N_XN_YN_Z(N_T+1)\) elements from the previous one (the
factor \(\frac{1}{2}\) arises because the fields are single parity).
The performance is now relatively constant, except at the smallest
volumes where the total number of threads is limited by the volume.

The benefit of using the 8 parameter reconstruction over that of the
12 parameter method can clearly be seen; the extra operations involved
for the 8 parameter method is not noticeable since the kernel is so
bandwidth starved.  The temporal gauge fixing brings additional
performance to both of these strategies, and the peak performance of
the 8 parameter kernel is 135 Gflops.

It is worth noting that performance of the GPU code is around an order
of magnitude greater than typical SSE-optimized implementations (which
generally achieve around 10 Gflops for Wilson-Dirac matrix-vector on Intel's
Nehalem architecture when operating in parallel~\cite{Holmgren:2009}).
In addition, the scaling of performance with volume is generally
reversed on CPU implementations which suffer dramatically as the local
volume falls out of cache.

\subsection{Double Precision Performance}
\label{sec:double}

For double precision our kernels translate almost directly, replacing
floats with doubles. However, both the gauge and spinor fields are
stored in arrays using the double2 primitive, since using double4
would destroy the coalescing in the memory reads.\footnote{To achieve
  coalescing, each thread must read data in either 32, 64, or 128 bit
  chunks.}  Although the peak double precision performance achievable
on current hardware is 12 times smaller than that of single precision,
the flops to bandwidth ratio is much more balanced in this case, and
so the difference in realized performance is expected to be smaller.
The memory traffic is doubled when going to double precision, so we
can expect at best 50\% of single precision performance.  Another
limiting factor is the register pressure caused by halving the number
of available registers, which limits occupancy for the kernel to
128 active threads.

\begin{figure}[htb]
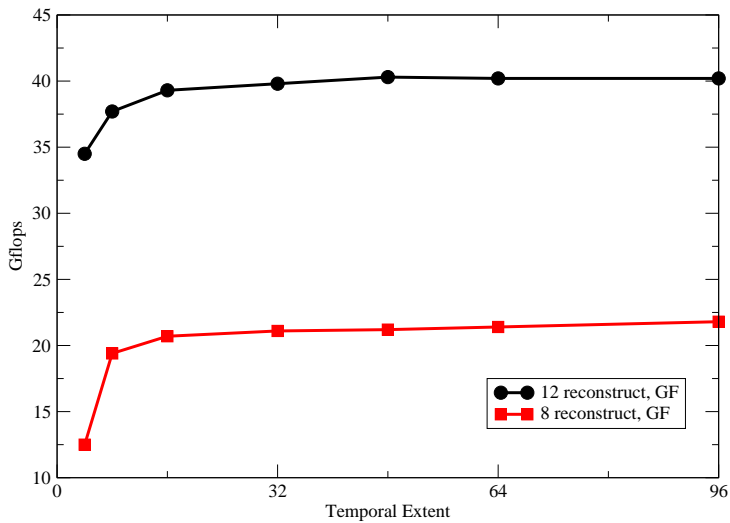

\begin{center}
\epspdffile{24_double}
\caption{\label{fig:wilson-double-flops}Performance of the double
  precision even-odd preconditioned Wilson-Dirac matrix-vector product
  on a GTX 280 as a function of temporal length (spatial volume
  \(24^3\); GF denotes that temporal gauge fixing has been applied).}
\end{center}
\end{figure}

In Figure \ref{fig:wilson-double-flops} the double precision
performance results are shown.  Here it is clear that the extra work
involved in using the 8-number parametrization of the gauge field
results in an overall decrease in performance relative to the
12-number parametrization.  The peak performance of the 12 parameter
kernel with gauge fixing is 40 Gflops, which is close to a factor of
three reduction relative to the equivalent single precision kernel.

\subsection{Half Precision Performance}
\label{sec:half}

Even with the modifications described in \S\ref{sec:bandwidth}
implemented, the single precision kernel is still bandwidth bound.
One way to decrease memory traffic further is to truncate the
precision of the gauge field elements and/or the spinor field
elements.

Although 16-bit floating point arithmetic has long been supported by
GPUs, until recently it was not supported by the CUDA runtime
API.\footnote{Intrinsics for conversion between 16-bit floats and
  32-bit floats were introduced in CUDA 2.3.  We find that kernel
  variants using these intrinsics are in certain scenarios faster, but
  are less numerically stable than the approach described here,
  depending on such factors as the type of reconstruction used for the
  gauge links and the condition number of the matrix.} There is,
however, a texture read mode called {\it cudaReadModeNormalizedFloat}.
When a texture is defined using this mode, a signed 16-bit (or even
8-bit) integer read in from device memory will be automatically
converted to a 32-bit floating point number in the range \([-1,1]\).
Clearly, such a fixed point format is well suited to storing the gauge
field elements since they all lie in this range.\footnote{The
  exception here being the first two components of the packed 8
  parameter format, which lie in the range \([-\pi,\pi]\), hence an
  extra multiplication by \(\pi\) is required.}  Using this format for
the spinor field elements is less straightforward since these elements
have no bounds, and the magnitudes of these elements can vary
drastically over the lattice.  However, because locally all color and
spin components are mixed together when multiplied by the gauge field
and spin projectors, a local normalization (exponent) of each
color-spinor in spacetime can be justified.  Thus we represent each
color-spinor by \(24\times\)16-bit integers (stored using
\(6\times\)short4 fields), together with a 32-bit float that
represents the maximum element of that color-spinor.  This
modification reduces the memory traffic by almost a factor of 2, and
so we would expect a large increase in the performance.

\begin{figure}[htb]
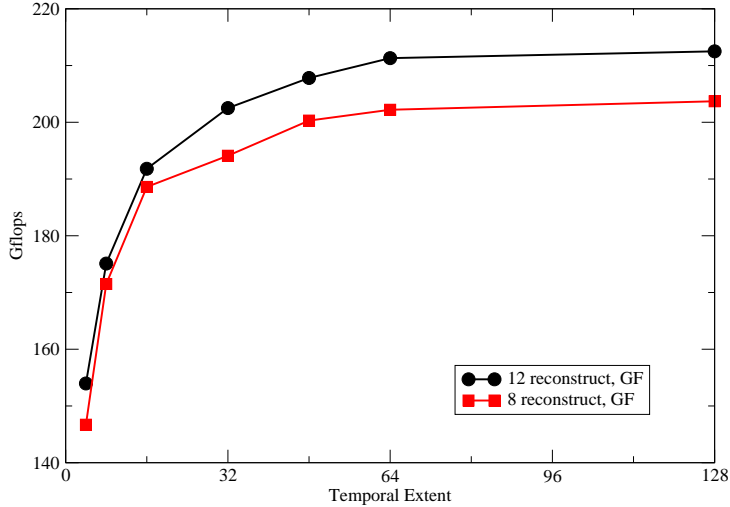

\begin{center}
\epspdffile{24_half}
 \caption{\label{fig:wilson-half-flops}Performance of the half
   precision even-odd preconditioned Wilson-Dirac matrix-vector
   product on a GTX 280 as a function of temporal length (spatial
   volume \(24^3\); GF denotes that temporal gauge fixing has been
   applied).}
\end{center}
\end{figure}

The performance of the pseudo half precision kernels can be seen in
Figure \ref{fig:wilson-half-flops}.  As was the case for double
precision, the 12 parametrization is faster than the 8
parametrization.  The additional overhead of unpacking and packing the
spinor fields together with reconstructing the gauge field is
equivalent to a doubling of cost of the matrix-vector product, and
thus the performance is finally floating point limited.  The peak
performance of the 12 parameter kernel is 212 Gflops which is 50\%
faster than the fastest single precision kernel.

\subsection{Performance versus accuracy}
\label{sec:accuracy}

\begin{figure}[htb]
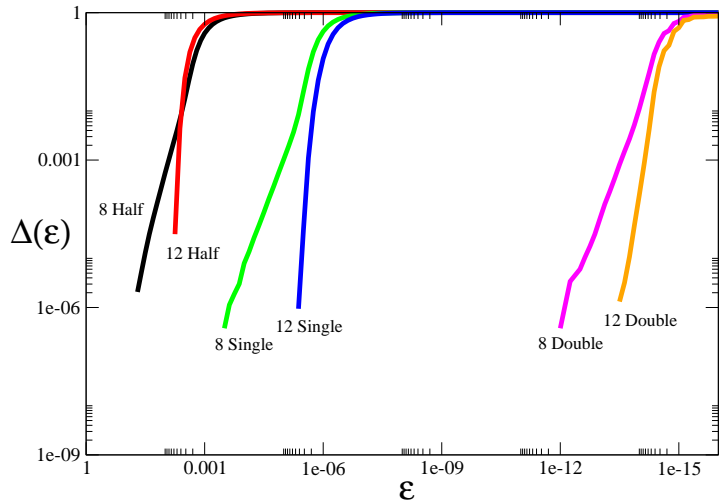

\begin{center}
\epspdffile{error}
\caption{\label{fig:wilson-error}Proportion \(\Delta\) deviation of
  GPU kernels from double precision CPU kernel as a function of
  varying tolerance \(\varepsilon\) (cf.~Equation \ref{eq:Delta}; volume =
  \(24^3\times 32\)).}
\label{fig:error}
\end{center}
\end{figure}

It is of course critical to investigate the numerical accuracy and
stability of the kernels described above.  To gain insight into this
we compared the element by element difference between the output of a
double precision CPU kernel and that of each of the GPU kernels
described above.  In particular, in figure \ref{fig:wilson-error} we
plot the proportion \(\Delta\) of the GPU output vector's components
that deviate from the CPU calculation as a function of varying
tolerance \(\varepsilon\), i.e.,
\begin{equation}
\Delta(\varepsilon) = \frac{1}{N} \sum_i \delta_i(\varepsilon),
\label{eq:Delta}
\end{equation}
with
\[
\delta_i(\varepsilon) = 
\left\{ \begin{array}{ll}
    1 & \mbox{if} \hspace{1mm} ||\sum_j(\hat{M}^{GPU}_{e,e})_{i,j} \eta_j - \sum_j(\hat{M}^{CPU}_{e,e})_{i,j} \eta_j|| \ge \varepsilon \\
        0 & \mbox{otherwise}\end{array} \right. ,
\]
and where \(N\) is the total number of degrees of freedom, the indices
\(i,j\) represent all degrees of freedom, \(\eta\) is a random vector
with elements \(\in[0,1]\) and random \(SU(3)\) matrices are used for
the gauge field.  Here, \(\hat{M}^{GPU}_{e,e} \) represents the
transfer of the CPU vector to the GPU, the application of the matrix
to the vector, and transfer back to the CPU.

Since both the vector and gauge field elements are random in this
test, and given that the Dirac matrix has a uniform sparsity pattern,
Equation \ref{eq:Delta} represents a stochastic estimate of the
cumulative distribution function of the accuracy of the different GPU
kernels.
The left most point of each curve is the first point at which the CPU
and GPU measurably deviate; hence \(\varepsilon\) at this point can be
interpreted as the accuracy of the corresponding kernel.  For all
three different precisions, the 12 parameter reconstruction is the
most numerically stable, with at least one to two orders of magnitude
increased accuracy over the 8 parameter reconstruction.  This is
because the latter involves many more operations, including
trigonometric functions and, crucially, division by a subtracted
quantity.  For both single and double precisions, the 12 parameter
method exhibits deviations at a value of \(\varepsilon\) around an
order of magnitude greater than the respective unit of least precision
(ulp).


\section{Mixed Precision Krylov Solvers}
\label{sec:inverter}

\subsection{Implementation and Performance}

We have implemented both conjugate gradients (CG) and BiCGstab solvers
on the GPU for all three of the precisions considered above.  Since
the Wilson-Dirac matrix is not Hermitian positive definite, CG is
applied using the normal equations, and BiCGstab is applied to the
problem directly.  Since Krylov solvers can be decomposed into simple
linear algebra operations, e.g., global sums, scaling and adding
vectors (AXPYs) and of course the matrix-vector product, it is natural
to implement each of these operations as a separate CUDA kernel.
However, since all of these operations are extremely bandwidth bound,
where possible we have fused these operations to minimize memory
traffic; e.g., in CG the AXPY operation to update the residual and the
calculation of its norm can be combined into a single kernel.  The
fact that CUDA supports C++ style templates drastically reduced the
development time for these kernels for the different precisions,
though the half precision kernels required additional attention since
the vector elements must be read in blocks of 24 numbers and double
precision complex valued kernels greatly benefited from using the
texture cache.

For reliable convergence the global sums must be done as accurately as
possible.  On CPU implementations, this typically means that,
regardless of the precision of the matrix-vector product, the global
sum accumulators are double precision variables.  On first generation
CUDA devices this poses a problem since double precision is not
implemented, so schemes such as Kahan summation~\cite{Kahan:1965} are
required to reduce the accumulation of errors.  When running on second
generation devices this restriction does not apply, so true double
precision reductions are always used.  For such reductions we follow
the approach advocated in \cite{Harris} which uses tree-based
reduction in shared memory, split over several kernel invocations to
achieve complete reduction over the lattice.

\begin{table}[htb]
\begin{center}
\begin{tabular}{|l|c|c|c|}\hline 
Kernel type & Kernel (Gflops)& CG (Gflops) & BiCGstab (Gflops) \\ \hline 
12 Half GF & 207.5 & 179.8 & 171.1 \\ \hline 
8 Single GF & 134.1 & 116.1 & 109.9 \\ \hline 
12 Single GF & 122.1 & 107.7 & 102.6 \\ \hline 
12 Double GF & 40.3 & 38.3 & 37.9 \\ \hline
\end{tabular}
\caption{\label{table:invert-perf}Performance comparison of the
 matrix-vector kernels with the associated CG and BiCGstab solvers
 on the GeForce GTX 280 (volume = \(24^3\times48\)).}
\end{center}
\end{table}
  
A performance comparison of the solvers and matrix-vector kernels are
given in Table \ref{table:invert-perf}.  In both half and single 
precision, CG and BiCGstab run at around 85\% of the associated
matrix-vector kernel because of the additional memory bandwidth
intensive linear algebra operations.  In double precision, this is
less pronounced since here the GTX 280 is flop limited.  As shall be
shown in \S \ref{sec:results}, these measurements are not necessarily
a good measure of solver performance, since the only metric that we
care about is time to an accurate solution.

\subsection{Defect-Correction}

The simplest approach when implementing a mixed precision solver is to
use an iterative refinement strategy, also known as defect-correction,
shown in Algorithm \ref{alg:defect}~\cite{Wilkinson:1966}.  Such an
approach allows the residual to be reduced in an inner solve using low
precision, while the residual calculation and solution accumulation are
done in high precision.  This ensures that the method converges to the
desired precision \(\epsilon\) provided that the inverse of the
spectral radius is bounded by the unit of least precision of the
arithmetic used for the inner solve, i.e., \(1/\rho(A) >
\mbox{ulp}^{in}\).  However, when a Krylov solver is used for the
inner solve, each new solve results in the previously generated Krylov
sub-space being discarded: this can drastically increase the total
number of iterations of the solver to reach convergence.  The
parameter \(\epsilon^{in}\) is typically chosen to be as small as
possible to avoid unnecessary restarting, subject to the constraint
\(\epsilon^{in} > \mbox{ulp}^{in}\).

\begin{algorithm}[H]
\SetLine
\(r_0\) = \(b - Ax_0\)\;
\(k\) = 0\;
\While{\(||r_{k}|| > \epsilon\)}{
\mbox{Solve} \(A p_{k+1}\) = \(r_{k}\) \mbox{to precision} \(\epsilon^{in}\)\; 
\(x_{k+1}\) = \(x_k + p_{k+1}\)\;
\(r_{k+1}\) = \(b - Ax_{k+1}\) \;
\(k = k + 1\)\;
}
\label{alg:defect}
\caption{Defect-correction solver for \(Ax=b\) (initial guess
  \(x_0\), outer solver tolerance \(\epsilon\) and inner solver
  tolerance \(\epsilon^{in}\)).}
\end{algorithm}

\subsection{Reliable Updates}

Using low precision arithmetic in a Krylov solver will cause the
iterated residual to drift away from the true residual as the solver
proceeds.  Residual drift and possible cures have been studied
previously in different contexts \cite{Sleijpen:1996}, namely where
the drift is caused by the erratic convergence of BiCGstab which
induces rounding errors.  The cure advocated in \cite{Sleijpen:1996}
is that of reliable updates: here a parameter \(\delta\) is
introduced, and if the magnitude of the iterated residual decreases by
\(\delta\) compared to the magnitude of all previous residuals, the
iterated residual is replaced by the true residual.  Erratic
convergence also introduces a source of error into the iterated
solution because of irregular summation of the solution vector.  This
is solved by performing groupwise updates of this vector: if the
residual decreases by \(\delta\) the solution is summed to a separate
accumulator, the solution is set to zero and the source vector is set
to the residual.  Reliable updates allow precision to be maintained
without the overhead of restarting.

\begin{algorithm}[H]
 \SetKwData{True}{true}
\SetLine
\(r_0\) = \(b - Ax_0\)\;
\(\hat{r}_{0}\) = \(r\)\;
\(\hat{x}_{0}\) = 0\;
\(k\) = 0\;
\While{\(||\hat{r}_{k}|| > \epsilon\)}{

Low precision solver iteration: \(\hat{r}_{k}\rightarrow \hat{r}_{k+1}\), \(\hat{x}_{k}\rightarrow \hat{x}_{k+1}\)\;

\If{\(||\hat{r}_{k+1}||<\delta M(\hat{r})\)}{
\(x_{l+1}\) = \(x_{l}+\hat{x}_{k+1}\)\;
\(r_{l+1}\) = \(b - Ax_{l+1}\) \;
\(\hat{x}_{k+1}\) = 0\; 
\(\hat{r}_{k+1}\) = \(r\)\;
\(l\) = \(l+1\)\;
}
\(k\) = \(k + 1\) \;
}
\label{alg:reliable}
\caption{Reliable update solver for \(Ax=b\) (initial guess
  \(x_0\), outer solver tolerance \(\epsilon\), \(M(r)\) is the maximum of the norm of
  the residuals since the last residual update, ( \(\hat{}\) ) denotes low
  precision).}
\end{algorithm}

It should now be clear that reliable updates offer an alternative to
using defect-correction in the context of mixed precision solvers: if
the iterated residual and the solution vector are updated using a low
precision matrix-vector product, then any errors introduced can be
rectified periodically by using reliable updates in high precision.
The reliable update scheme we have adopted is shown in Algorithm
\ref{alg:reliable}.\footnote{For CG, one also has to take care when
  updating the gradient vector when a reliable update is
  performed~\cite{StGo06PipeCG}.}  Here we have simplified the approach
given in \cite{Sleijpen:1996} such that we perform a reliable residual
update whenever the norm of the residual decreases by a factor
\(\delta\) relative to the maximum of the residual since the last
update.  In addition, by always incorporating a groupwise solution
update whenever a residual update is done, the total number of high
precision spinor fields required can be reduced to 4 (source \(b\),
residual \(r\), solution \(x\) and an additional temporary required
when applying the even-odd Wilson-Dirac matrix).  Thus the total memory
overhead of using a reliable update scheme is the number of spinor
fields required for the low precision solver (5 for CG and 7 for
BiCGstab) in addition to 4 high precision spinor fields, as well as
the gauge field storage for both precisions.

The parameter \(\delta\) should be chosen such that \(\mbox{ulp}^{in}
< \delta < 1\) since any reduction beyond \(\mbox{ulp}^{in}\) will be
erroneous, causing algorithmic degradation, while on the other hand
\(\delta \sim 1\) is equivalent to an outer precision solve which will
cause raw performance degradation.

\subsection{Results}
\label{sec:results}

This work represents an important first step in exploring mixed
precision solvers on GPUs.  We present results that represent an
average over runs using five \(V = 24^3\times 64\) anisotropic
lattices,\footnote{This is the largest volume possible using double
  precision on a 1 GiB GPU.  These gauge fields were provided by the
  Hadron Spectrum Colaboration~\cite{Bulava:2009jb}, generated using
  \(\beta = 5.5\), \(m=-0.4125\), and an aspect ratio of 3 between the
  spatial and temporal lattice spacings.} using random sources.  We
note, however, that we have tested these methods on a range of
different lattice volumes and gauge couplings, and the conclusions we
draw are the same in all cases.

To provide a strong test of the mixed precision precision methods
described above, we set the desired final residual tolerance at
\(||r||<\epsilon=10^{-12}\), which is far beyond the limit of single
precision.  The quark mass parameter \(m\) was varied in the physical
range of interest, namely \([-0.4180,-0.3980]\), between the critical
mass (extremely light) and the approximate strange quark mass (heavy).
For a baseline comparison, Table \ref{table:invert-double} gives the
total number of iterations for the pure double precision CG and
BiCGstab solvers for this range of masses.  For brevity we have
omitted the sample errors, which are small and have no bearing on our
conclusions.  Using BiCGstab over CG results in an approximate
threefold decrease in iterations, and therefore also time to solution.
Thus, in the mixed precision results that follow, we concentrate on
BiCGstab.\footnote{We observe that our conclusions also hold for mixed
  precision CG for this problem and should generalize to problems
  where BiCGstab is unsuitable (e.g., inversion of the domain wall
  matrix).}

\begin{table}[htb]
\begin{center}
\begin{tabular}{|c|c|c|}\hline
 \(m\) &  CG & BiCGstab  \\ \hline
-0.3980 & 1392 & 460  \\ \hline
-0.4005 & 1570 & 509  \\ \hline
-0.4030 & 1794 & 570  \\ \hline
-0.4055 & 2079 & 666  \\ \hline
-0.4080 & 2698 & 785  \\ \hline
-0.4105 & 3226 & 939  \\ \hline
-0.4130 & 4055 & 1193 \\ \hline
-0.4155 & 5046 & 1545 \\ \hline
-0.4180 & 6734 & 2078 \\ \hline
\end{tabular}
\caption{\label{table:invert-double}Number of iterations until
  convergence of full double precision CG and BiCGstab solvers
  (\(\epsilon=10^{-12}\), volume = \(24^3\times64\)).}
\end{center}
\end{table}

\begin{table}[htb]
\begin{center}
\begin{tabular}{|l|c|c|c|c|c|c|c|}\hline
&  \multicolumn{2}{|c|}{Half 12} & \multicolumn{3}{|c|}{Single 8}  & \multicolumn{2}{|c|} {Single 12} \\ \hline
\backslashbox{\(m\)}{\(\epsilon^{in}\)} &  \(10^{-1}\) &  \(10^{-2}\) &  \(10^{-3}\) &  \(10^{-4}\) & \(10^{-5}\) & \(10^{-5}\) & \(10^{-6}\)\\ \hline
-0.3980 & 489 & 572 & 446 & 470 & 486 & 479 & 498 \\ \hline
-0.4005 & 543 & 632 & 477 & 541 & 563 & 535 & 561 \\ \hline
-0.4030 & 621 & 684 & 537 & 642 & 631 & 625 & 639 \\ \hline
-0.4055 & 783 & 871 & 637 & 702 & 753 & 753 & 722 \\ \hline
-0.4080 & 938 & 1055 & 738 & 816 & 861 & 878 & 914 \\ \hline
-0.4105 & 1152 & 1314 & 902 & 1042 & 1066 & 1013 & 1092 \\ \hline
-0.4130 & 1638 & 1870 & 1174 & 1262 & 1454 & 1341 & 1412 \\ \hline
-0.4155 & 2161 & 2581 & 1678 & 1699 & 1795 & 1830 & 1966 \\ \hline
-0.4180 & * & * & 2526 & 2523 & 2723 & 2738 & 2986 \\ \hline
\end{tabular}
\caption{\label{table:invert-defect}Number of iterations (including
  restart iterations) until convergence of defect-correction BiCGstab
  solver (double precision correction, \(\epsilon=10^{-12}\), volume =
  \(24^3\times64\), * = did not converge ).}
\end{center}
\end{table}

In Table \ref{table:invert-defect} we present results for
defect-correction where we have used the results from \S
\ref{sec:accuracy} to guide our choice of \(\epsilon^{in}\).  At heavy
quark mass it makes very little difference which inner precision is
used, nor is there much dependence on \(\epsilon^{in}\).  The numbers
of iterations until convergence are comparable to the full double
precision solver, and in some cases are even smaller; this we
attribute to the sensitivity of Krylov solvers in finite precision
arithmetic.  In this regime half precision is the method of choice.
At light quark masses the story changes. In particular as the mass is
reduced the half precision solver iteration count increases relative
to the double precision baseline, and completely fails to converge at
the critical mass, where likely the inverse spectral radius of the
matrix is greater than the resolution of half precision.  There is
little to choose between the two single precision kernel variants, and
it would appear that the penalty from frequent restarting (large
\(\epsilon^{in}\)) is less severe than that of the accumulation of
rounding errors from running the inner solver for longer (small
\(\epsilon^{in}\)).  In terms of iteration counts, the penalty for
using defect-correction at light quark masses ranges between 10 and
30\%.

\begin{table}[htb]
\begin{center}
\begin{tabular}{|l|c|c|c|c|c|c|c|c|}\hline
 &  \multicolumn{3}{|c|}{Half 12} & \multicolumn{3}{|c|}{Single 8}  & \multicolumn{2}{|c|}{Single 12} \\ \hline
\backslashbox{\(m\)}{\(\delta\)} &  \(10^{-1}\) &  \(10^{-2}\) &  \(10^{-3}\) &  \(10^{-1}\) &  \(10^{-2}\) & \(10^{-3}\) & \(10^{-2}\) & \(10^{-3}\) \\ \hline
-0.3980 & 490 & 509 & 556 & 461 & 467 & 470 & 465 & 469 \\ \hline
-0.4005 & 531 & 563 & 609 & 535 & 533 & 537 & 538 & 533 \\ \hline
-0.4030 & 634 & 706 & 737 & 610 & 605 & 616 & 580 & 583 \\ \hline
-0.4055 & 700 & 735 & 772 & 688 & 676 & 686 & 682 & 694 \\ \hline
-0.4080 & 856 & 856 & 973 & 795 & 809 & 821 & 796 & 807 \\ \hline
-0.4105 & 1060 & 1101 & 1273  & 950 & 958 & 1008 & 951 & 990  \\ \hline
-0.4130 & 1354 & 1484 & 1570 & 1176 & 1207 & 1268 & 1197 & 1225 \\ \hline
-0.4155 & 1860 & 2249 & 2146 & 1636 & 1595 & 1668 & 1584 & 1595 \\ \hline
-0.4180 & 2793 & 3483 & 3319 & 2147 & 2369 &  2191 & 2256 & 2159 \\ \hline
\end{tabular}
\caption{\label{table:invert-reliable}Number of iterations (including
  reliable updates, cf. \(k+l\) from Algorithm \ref{alg:reliable})
  until convergence of reliable update BiCGstab solver (double
  precision reliable update, \(\epsilon=10^{-12}\), volume =
  \(24^3\times64\)).}
\end{center}
\end{table}

In Table \ref{table:invert-reliable} we present results for the
reliable update aproach.  As was the case for defect-correction, at
heavy quark masses there is little to choose between the methods and
their respective precisions and parameters.  For lighter quark masses,
again the iteration counts increase relative the double precision
equivalent, but now the increase is much milder.  As expected, the
general trend is that larger values of \(\delta\) correspond to
iteration counts closer to the double precision baseline.  For single
precision, the increase in iterations at light quark masses is never
more than 15\%, and now half precision converges at the lightest mass, with
an increase of 34\% for \(\delta=10^{-1}\).

The decision on whether to use defect-correction or reliable updates
is a pragmatic one.  At heavy quark masses, there is little
difference, but defect-correction has the advantage that the
correction can be staged on the CPU because of the infrequent
corrections.  Thus for problems that are memory constrained,
defect-correction can be the better choice.  However, given the poor
performance of defect-correction at light quark masses, reliable
updates is the method of choice if memory is not a concern.

\begin{figure}[htb]
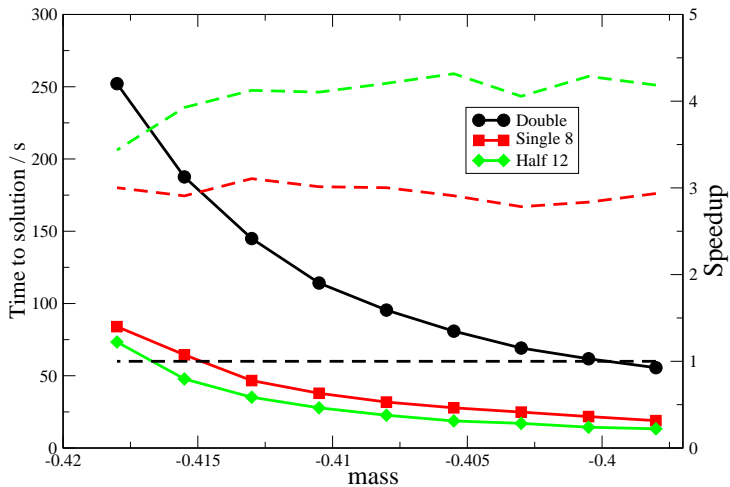

\begin{center}
\epspdffile{time}
\caption{\label{fig:time} Time to solution (solid lines) for double
  precision and reliable update solvers and speedup versus double
  precision (dashed lines) (BiCGstab, \(\delta=0.1\),
  \(\epsilon=10^{-12}\), volume = \(24^3\times 64\)).}
\end{center}
\end{figure}

The metric of real interest is of course time to solution, or the
speedup relative to the double precision solver.  In Figure
\ref{fig:time} we plot the time to solution of the double precision
solver and two of the reliable update solvers (Half 12 and Single 8 at
\(\delta=10^{-1}\)), as well as their relative speedups versus double
precision.  Here it is plain to see that the small overhead in the
increase of iterations is more than made up for by faster performance.
The Single 8 solver has an almost constant factor 3 speedup, while the
Half 12 solver varies from a factor of 4 to 3.5 as the quark mass is
reduced.  Given the performance of the double precision solver (\(\sim
39\) Gflops), this corresponds to an effective solver performance well
in excess of 100 Gflops.

\section{Conclusions}
\label{sec:conclusion}

In summary we have implemented optimized Wilson-Dirac matrix-vector
routines, upon which we have constructed Krylov solvers on the CUDA
architecture.  The key to obtaining high performance lies in
minimizing the memory bandwidth requirements even at the expense of
increasing the overall floating point operation count.

We have introduced an alternative mixed precision method, using
reliable updates, which we compared against the conventional
defect-correction approach.  For the systems tested, we found the
combination of a half precision BiCGstab solver together with double
precision reliable updates to give the least time to solution.

At 100 Gflops sustained inverter performance GPUs represent a
considerable cost and power savings over traditional cluster and
massively parallel architectures.  As new GPU solutions come to
market, we expect GPUs to become even more attractive for QCD
calculations.

The linear solver developed in this work has become the mainstay of
our open source {\it QUDA} library~\cite{quda}, which we have
interfaced to the common lattice QCD packages
(Chroma~\cite{Edwards:2004sx, chroma}, CPS~\cite{cps},
QDP/C~\cite{qdp}) for easy integration with current QCD calculations.
As well as the Wilson-Dirac matrix, for which results are reported
here, the library currently also supports the Sheikholeslami-Wohlert
(also known as Wilson-Clover) discretization of the Dirac operator.

Future work in this area shall concentrate on implementing other
computationally costly operations needed for lattice QCD on the CUDA
architecture.  We shall also be developing an appropriate variant of our
adaptive multigrid solver~\cite{Brannick:2007ue, Clark:2008nh}; here
we expect the combination of a superior linear solver with the cost
performance of GPUs to be an excellent combination.  Adapting our code
to work in a multi-GPU environment is already underway, the
performance of which will be reported in future work.

This work was supported in part by US DOE grants DE-FG02-91ER40676 and
DE-FC02-06ER41440 and NSF grants DGE-0221680, PHY-0427646, PHY-0835713
and OCI-0749300.  We would like to thank D.~Luebke of {\nvidia} for
generous hardware donations and G.~Shi for improving the numerical
stability of the half precision 8 parameter reconstruction.

\appendix
\section{\(SU(3)\) Matrix Reconstruction}
\label{app:su3}
Label the components of a general \(SU(3)\) matrix as follows
\[
\left(\begin{array}{c}
    {\bf a}  \\
    {\bf b} \\
    {\bf c}
 \end{array}\right)
=
\left(\begin{array}{ccc}
    a_1 & a_2 & a_3 \\
    b_1 & b_2 & b_3 \\
    c_1 & c_2 & c_3
 \end{array}\right).
\]

\subsection{12 number parametrization}
If only the first 2 rows are stored, the third row is given by
\cite{De Forcrand:1986af}
\[
  {\bf c} = ({\bf a} \times {\bf b})^*. 
\]

\subsection{8 number parametrization}
While it would be possible to explicitly store the eight $SU(3)$
generators and reconstruct the link matrices on the fly, this would be
too costly in terms of operation count.  We have modified the method
described in \cite{Bunk:1985rg} as follows.

\begin{enumerate}
\item Given the row vector \({\bf a}\), the vectors
  \begin{eqnarray*}
    {\bf b}' &  = & \frac{1}{N} (0, -a_3^*, a_2^*) \\
    {\bf c}' & = & ({\bf a}\times{\bf b'})^*\\
    & = & (N, -\frac{1}{N}a_1^*a_2, -\frac{1}{N}a_1^*a_3),
\end{eqnarray*}
with \(N = \sqrt{|a_2|^2+|a_3|^2}\), define a plane orthogonal to \({\bf a}\).
\item Since the row vectors \({\bf b}\) and \({\bf c}\) must lie in
  this plane, the original \(SU(3)\) matrix can be obtained from an
  \(SU(2)\) rotation of this plane,
\[
\left(\begin{array}{ccc}
    a_1 & a_2 & a_3 \\
    b_1 & b_2 & b_3 \\
    c_1 & c_2 & c_3
 \end{array}\right)
=
\left(\begin{array}{ccc}
    1 & 0 & 0 \\
    0 & p_1 & p_2 \\
    0 & -p_2^* & p_1^*
 \end{array}\right)
\left(\begin{array}{ccc}
    a_1 & a_2 & a_3 \\
    0 & -\frac{1}{N}a_3^* & -\frac{1}{N}a_2^* \\
    N & -\frac{1}{N}a_1^*a_2 & -\frac{1}{N}a_1^*a_3
 \end{array}\right).
\]
Equating right and left matrix elements we immediately see that \(b_1
= Np_2\) and \(c_1^* = Np_1\), hence the matrix may be parametrized
by \(a_1, a_2, a_3, b_1, c_1\), i.e., 10 real components.
\item Finally we use the normality of the first row and column to
  reduce to 8 real numbers.  In this step, \cite{Bunk:1985rg} favored a
  stereographic projection approach. However, this introduces two
  singularities in the reconstruction.  Instead we favor storing the
  phases of \(a_1\) and \(c_1\), obtaining the full number through
  trigonometric functions,
\begin{eqnarray*}
  a_1 & = & \sqrt{1 - |a_2|^2 - |a_3|^2} (\cos \theta_{a_1} +i\sin\theta_{a_1} )\\
  c_1 & = & \sqrt{1 - |a_1|^2 - |b_1|^2} (\cos \theta_{c_1} +i\sin\theta_{c_1} ).
\end{eqnarray*}
This is much more numerically stable and can be evaluated very
efficiently on GPU architectures because of the presence of fast
trigonometric and square root functions.  In half precision the
argument to the square root can be negative, in which case setting
it to zero vastly improves the numerical stability.  We note, however,
that this reconstruction still has a singularity at \(|a_1| = 1\) due
to the normalization factor \(N\).\end{enumerate}

\section{Dirac Matrix Conventions}
\label{app:gamma}

It is somewhat conventional in lattice QCD software to use a chiral
basis for the spin projectors that appear in the off-diagonals of the
Wilson-Dirac matrix.  An example is the DeGrand-Rossi basis, in which
the projectors are given by
\[
P^{\pm 1} =
 \left(\begin{array}{rrrr}
   1&0&0&\pm i\\
   0&1&\pm i&0\\
   0&\mp i&1&0\\
   \mp i&0&0&1\\
 \end{array}\right),\,
P^{\pm 2} = 
 \left(\begin{array}{rrrr}
   1&0&0&\mp 1\\
   0&1&\pm 1&0\\
   0&\pm1&1&0\\
   \mp1&0&0&1\\
 \end{array}\right),\]
\[P^{\pm 3} = 
 \left(\begin{array}{rrrr}
   1&0&\pm i&0\\
   0&1&0&\mp i\\
   \mp i&0&1&0\\
   0&\pm i&0&1\\
 \end{array}\right),\,
P^{\pm 4} = 
 \left(\begin{array}{rrrr}
   1&0&\pm1&0\\
   0&1&0&\pm1\\
   \pm1&0&1&0\\
   0&\pm1&0&1\\
 \end{array}\right).
\]
Hence when applying this projection to a spinor we must always load
all components regardless of the dimension or direction.  An
alternative is the ``non-relativistic'' or UKQCD basis, in which the
projectors have the form
\[
P^{\pm 1} =
 \left(\begin{array}{rrrr}
   1&0&0&\pm i\\
   0&1&\pm i&0\\
   0&\mp i&1&0\\
   \mp i&0&0&1\\
 \end{array}\right),\,
P^{\pm 2} = 
 \left(\begin{array}{rrrr}
   1&0&0&\pm 1\\
   0&1&\mp 1&0\\
   0&\mp1&1&0\\
   \pm1&0&0&1\\
 \end{array}\right),\]
\[P^{\pm 3} = 
 \left(\begin{array}{rrrr}
   1&0&\pm i&0\\
   0&1&0&\mp i\\
   \mp i&0&1&0\\
   0&\pm i&0&1\\
 \end{array}\right),\,
P^{+4} = 
 \left(\begin{array}{rrrr}
   2&0&0&0\\
   0&2&0&0\\
   0&0&0&0\\
   0& 0&0&0\\
 \end{array}\right),\]
\[P^{-4} = 
 \left(\begin{array}{rrrr}
   0&0&0&0\\
   0&0&0&0\\
   0&0&2&0\\
   0& 0&0&2\\
 \end{array}\right).
\]
An advantage of using such a basis is that in the temporal direction
we need only load the upper (lower) spin components for the backwards
(forwards) gather.  This halves the memory traffic needed to perform
the temporal gather, and so increases the kernel's performance.

For easy interfacing with current lattice QCD packages that use the
DeGrand-Rossi basis, this basis transformation is applied whenever a
spinor field is transferred from host to device, and undone when
transferred from device to host.

\end{document}